**Engineered Bacteria Computationally Solve Chemically Generated 2X2 Maze Problems**

Kathakali Sarkar, Deepro Bonnerjee, Sangram Bagh[*]

Biophysics and Structural Genomics Division, Saha Institute of Nuclear Physics, Homi Bhabha National Institute (HBNI), Block A/F, Sector-I, Bidhannagar, Kolkata 700064 INDIA

[*] Corresponding author E-mail: sangram.bagh@saha.ac.in

**Abstract:** Maze generating and solving are challenging problems in mathematics and computing. Here we generated simple 2X2 maze problems applying four chemicals and created a set of engineered bacteria, which in a mixed population worked as a computational solver for any such problem. The input-output matrices of a mathematical maze were mapped through a truth table, where the logic values of four chemical inputs determined the sixteen different 2X2 maze problems on a chemical space. Our engineered bacteria, which consisted of six different genetic logic circuits and distributed among six cell populations processed the chemical information and solved the problems by expressing or not expressing four different fluorescent proteins. The three available 'solutions' were visualized by glowing bacteria and for the thirteen 'no solution' cases no bacteria glowed. Thus, our system not only solved the maze problems but also showed the number of solvable and unsolvable problems. This work presented a new and abstract cellular computational way towards maze problems and may have significance in biocomputation and synthetic biology.

Maze solving is a popular puzzle throughout human history. The problem is to search for an un-obstructed way from the start point to the destination point within a maze. However, maze generating and solving is difficult for lower animals and robots and it became a fundamental problem in mathematics and computation[1,2]. Apart from human and computer based solutions, mazes have been solved based on physical and chemical phenomena[3]. Some of the examples include maze solving using surface tension[4], glow discharge[5] Belousov–Zhabotinsky reaction[6], Marangoni flow[7], chemotactic droplet[8], and hybridization chain reaction[9].

Maze problems also have been solved by living biological cells including slime mould[10,11] and cancer cells[12]. All cell based maze solving were based on the same basic principle, where the problems were confined to physical maze structures and the solutions were depended on chemotaxis gradients. In those cell-based approaches, the problem generation and solution were independent or decoupled, thus, it cannot suggest the plausible number of solvable and unsolvable mazes of a specific size.

The generation of maze problems and its solutions can be abstracted as a mathematical or computational systems beyond any physical structure[13-15]. A common method of abstracting a maze problem of a particular size is to create a maze by a binary matrix with 1(s) and 0(s), where '0' represents obstruction and '1' represents open space. This data structure stores the neighbourhood information within the matrix. By changing the positions and the values (0 and 1) of the matrix of a specific size all pluasible mazes of that size can be geneareted and such problem space can be represented as a set of matrices. Any maze from those matrices can be solved by using some computational algorithm, where the solutions are represented as a sequence of moves. The solutions are again represented by a solution matrix, where '0' suggests no move and '1' suggests a move. The solution path can be created if and only if there is a path with all the spaces within the path have values 1.

The advent of synthetic biology has been demonstrated the adaptation of electronics engineering principles in the realm of cellular biology to build many electronic analogous synthetic genetic logic circuits for performing various computational operations[16-18]. Some of the examples of such devices include basic logic gates and integrated circuits[19-22], half adders[23], and DeMux and Mux [24]. The logical operations have been implemented in genetic devices by engineering the transcriptional and translational machinery in bacterial cells.Adaptation of such concepts inspires the new possibilities of implementing abstract computational maze solving with living engineered cells.

Here we demonstrated that a mixed population of engineered bacteria worked as a computational algorithm and solved any plausible 2X2 maze problem generated on a chemical space. All possible problems within 2X2 maze were generated by applying four small molecule chemicals and that chemical information were processed by a set of synthetic genetic circuits distributed among a mixed population of bacteria, which responds by expressing or not expressing fluorescence proteins. The 'solutions' were visualized through fluorescently glowing bacteria. When there was no solution, no bacteria glow.

**Principles of maze problems generation and solution:** First we represented the maze problem in an abstract mathematical way. We considered a 2X2 maze and placed start and stop points. When these two points were fixed, we could place the obstructions in sixteen different ways, such that three different solutions were possible and in the rest thirteen cases, there were no solutions (Fig. 1A). This scenario was abstracted with problem and solution matrices, where problem matrices posed the problem and solution matrices showed the solutions or its absence (Fig. 1A). In each problem matrix, the value '0' suggested the obstruction and the value '1' suggested the open space. Similarly, in a solution matrix, '0' suggested the 'no move' and '1' suggested the 'move'. All those individual problem and solution matrices together, gave a complete input-out relation map, which we considered as a truth table (Fig 1B). This input-output relation map or truth table could be solved computationally with an electronic analogous genetic circuit of the following design (Fig. 1C).

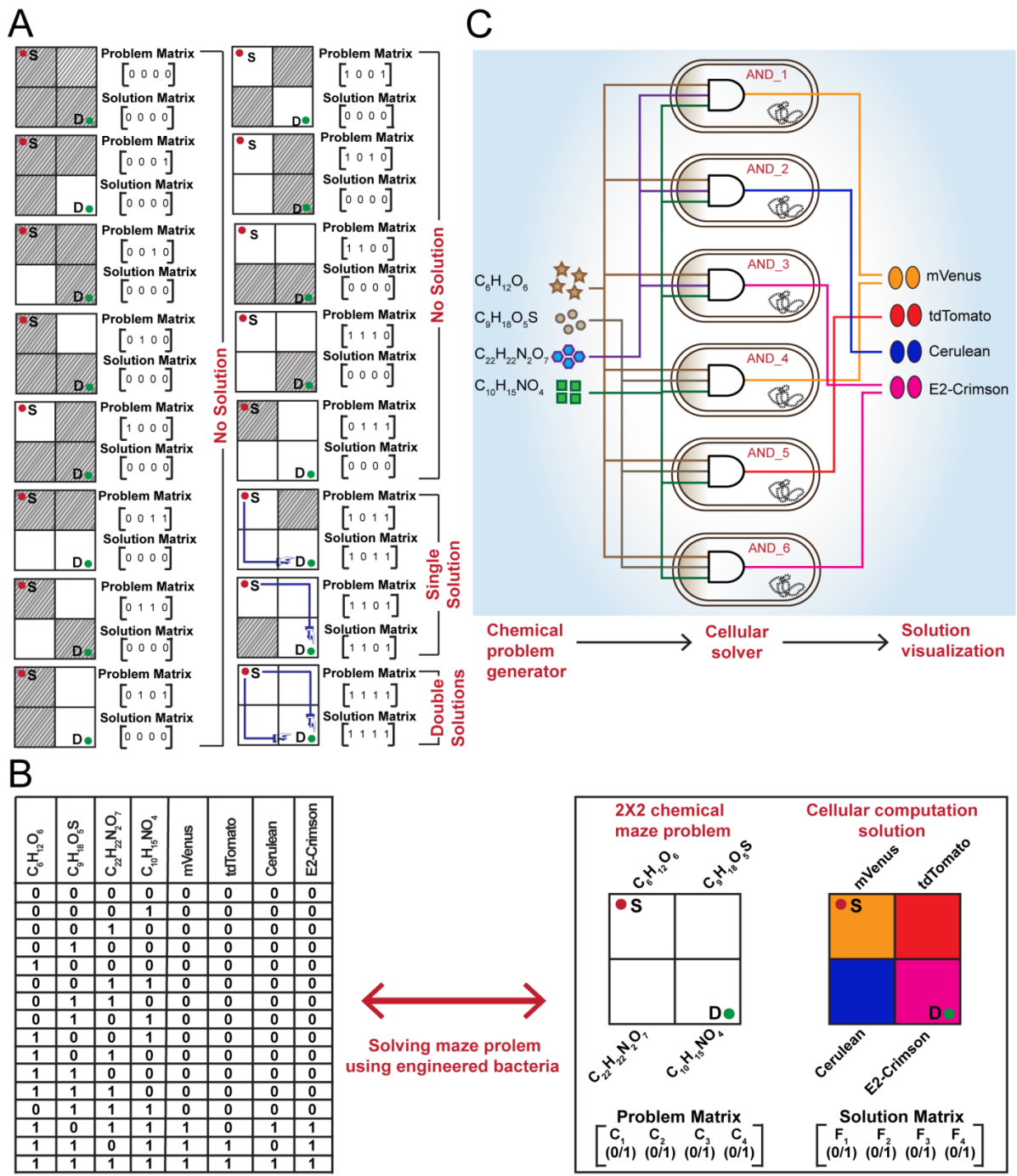

**Figure 1: Converting a 2X2 grid maze into a chemical maze problem and solving it with engineered bacteria. A)** All possible problems and corresponding solutions of a 2X2 grid maze are shown. The problems with no solution, single solution and double solutions are indicated. The blocked grids are shown with filled grey lines and the open grids are shown as white. Binary problem matrices were generated for individual mazes by considering the blocked (square filled with grey lines) and free grids (white square) as '0' and '1'

respectively. Solution matrices for each problem matrices were shown with valid path (1) or no path (0) **B)** The input-output relations obtained from the problem and solution matrices are organized in the form of a 4-input-4-output truth table where inputs are 4 chemicals ($C_6H_{12}O_6$ (D-glucose), $C_9H_{18}O_5S$ (IPTG), $C_{22}H_{22}N_2O_7$ (aTc) and $C_{10}H_{15}NO_4$ (3-oxo-C6-AHL)) and outputs are 4 fluorescent proteins (mVenus, tdTomato, Cerulean and E2-Crimson). The truth table is mapped as a problem generator and cellular computing solution (indicated with arrow). **C)** The design principle of the cellular solvers derived from the truth table. The mixed population of six bacteria carries six 3-input AND gate (AND_1-6).

Thus, any 2X2 maze problem could be created by turning ON (1) and OFF (0) the four input chemicals and the solution would be read by looking into the expression pattern of the four output fluorescent proteins, which would easily be visualized through fluorescence microscope (Fig. 1B,C).

We chose four different small molecule chemicals glucose ($C_6H_{12}O_6$), isopropyl β-D-1-thiogalactopyranoside ($C_9H_{18}O_5S$), anhydrotetracycline ($C_{22}H_{22}N_2O_7$), and N-Acyl homoserine lactones ($C_{12}H_{19}NO_4$), and designated them specific space in the 2X2 maze (Fig. 1B). Here, on a designated position on that maze, the presence and absence of a specific chemical were presented with the value '1' and '0' respectively. Thus, a maze problem was chemically represented. All possible set of problems (total 16) represented by a problem matrix and defined the problem space (Fig. 1A,B). Now, we created a set of synthetic genetic devices and inserted in bacteria, which in a mixed population worked as a maze solver (Fig 1C) by exactly following the input matrices-output matrices relation map (Fig. 1A,B). The genetic circuits processed those four chemical inputs and produce specific outputs by expressing (value '1') or not expressing (value '0') four fluorescence reporter proteins mVenus (yellow), TdTomato (orange), cerulean (cyan), and E-2 crimson (red) for designated positions on the 2X2 maze (Fig. 1B) and the solution could be visualized by glowing bacteria.

**Design, construction and initial characterization of the integrated molecular genetic circuit:**

In order to match the input-output relation map, we designed six multi-inputs genetic AND gates in *E.coli* cells, which in a mixed population worked as a maze solver (Fig. 1C). The basic molecular genetic design of these six cell sets was based on two basic genetic AND circuits, named as AND gate A and AND gate B (Fig. A and E).

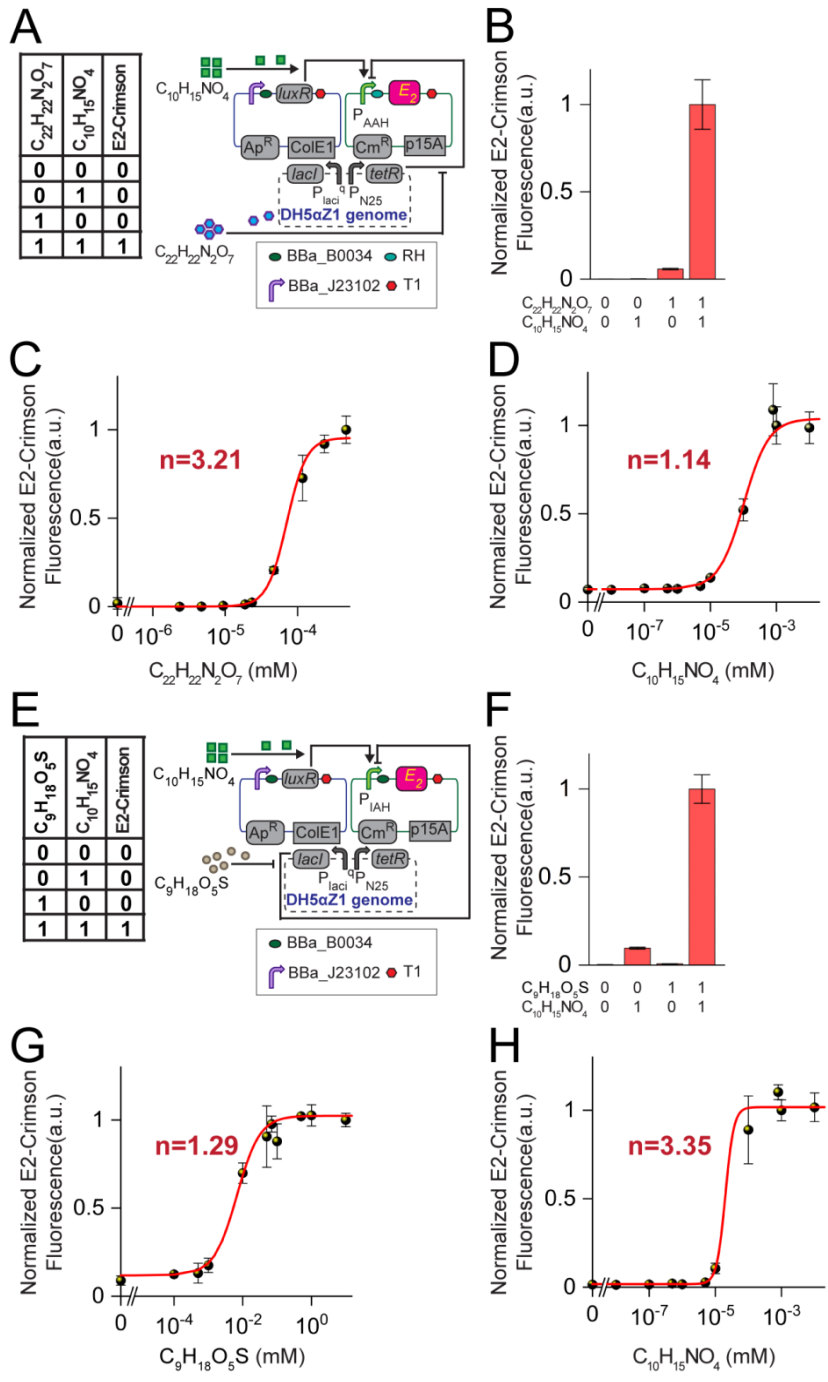

**Figure 2: Basic AND gates A and B.** Truth table and synthetic genetic circuit design for **A)** AND gate A and **E)** AND gate B. Experimental characterization of **B)** AND gate A and **F)** AND gate B. **C)** $C_{22}H_{22}N_2O_7$ and **D)** $C_{10}H_{15}NO_4$ dose responses for AND gate A. **G)** $C_9H_{18}O_5S$ and **H)** $C_{10}H_{15}NO_4$ dose responses for AND gate B.

The AND gate A (Fig. 2A) consisted of a synthetic promoter, $P_{AAH}$[25] (Table S1) having a lux box operating site upstream to the -35 position and two TetR binding sequences between -35 and -10 and downstream to the -10 position. The LuxR protein binds in the lux box operating sites. In the presence of $C_{12}H_{19}NO_4$, there is a change in conformation of the bound LuxR protein, which helps the proper binding of RNA polymerase in the promoter and start transcription[21]. The TetR protein binds in its operating sites as a dimer and obstructs the binding of the RNA polymerase. The small molecule $C_{22}H_{22}N_2O_7$, when applied from the outside, it diffuses into the cell, binds with the TetR and changes its conformation in such a way that it could no longer binds with the operating sites[26]. Thus, simultaneous presence of both $C_{22}H_{22}N_2O_7$ and $C_{12}H_{19}NO_4$ freed the promoter for RNA polymerase binding and activated for transcription. This gene circuit would work as an AND gate with respect to $C_{22}H_{22}N_2O_7$ and $C_{12}H_{19}NO_4$. In the AND gate B (Fig. 2E), we used another synthetic promoter, $P_{IAH}$[25] (Table S1), where instead of TetR binding sites, we used LacI binding sites between the TATA boxes and downstream of the -10 position. Here LacI tetramers bind in its operating sites and obstructs the binding of the RNA polymerase. The small molecule $C_9H_{18}O_5S$ binds with the tetramer and changes its conformation such that it could no longer bind with the operating sites[26]. Thus, the simultaneous presence of $C_{12}H_{19}NO_4$ and $C_9H_{18}O_5S$ activated the promoter and worked the second AND gate (AND gate 2). For both the cases TetR and LacI proteins were constitutively expressed from the engineered chromosome of *E.coli* strain DH5αZ1[26] and LuxR proteins were constitutively expressed from a plasmid under a J213 promoter[27] (Table S2). We used E2Crimson genes in both the basic AND gates as a fluorescent reporter for initial characterization. The experimental results showed that E2-crimson expressed in high amount only in the simultaneous presence of $C_{12}H_{19}NO_4$ and $C_{22}H_{22}N_2O_7$ and of $C_{12}H_{19}NO_4$ and $C_9H_{18}O_5S$, for AND gate A (Fig. 2B) and AND gate B (Fig. 2F) respectively.

To test, whether the output protein expression from both the synthetic genetic AND gates shows a digital like or ultrasensitive behavior as a function of the input variable, i.e. the concentration of inducer chemicals, we performed dose response experiments for both the AND gates. For the both AND gates, we measured the E2-Crimson florescence as a function of one inducer concentration, where the other was kept constant at its saturated concentration (Fig 2. C, D for AND gate A and Fig.2 G, H for AND gate B). We fitted those dose response curves with the simple transfer function models we developed.

The E2-crimson expression from the AND gates A and B at can be expressed by the equations 1 and 2 respectively,

$$\frac{d[E2Crimson]}{dt} = k\left(b_1 + \frac{[C_{22}H_{22}N_2O_7]^{n_1}}{K_1^{n_1} + [C_{22}H_{22}N_2O_7]^{n_1}}\right)\left(b_2 + \frac{[C_{10}H_{15}NO_4]^{n_2}}{K_2^{n_2} + [C_{10}H_{15}NO_4]^{n_2}}\right) - k_d[E2Crimson]$$

**equation 1**

$$\frac{d[E2Crimson]}{dt} = k'\left(b_3 + \frac{[C_9H_{18}O_5S]^{n_3}}{K_3^{n_3} + [C_9H_{18}O_5S]^{n_3}}\right)\left(b_4 + \frac{[C_{10}H_{15}NO_4]^{n_4}}{K_4^{n_4} + [C_{10}H_{15}NO_4]^{n_4}}\right) - k'_d[E2Crimson]$$

**equation 2**

Where, k,k' are the scaling rate constants, $k_d$, $k'_d$ are the rate of the E2-Crimson protein degradation, $b_i$ (s) are the basal level fluorescence, $K_i$(s) are the Hill constants, and $n_i$(s) are the Hill coefficients. At steady state, $\frac{d[E2Crimson]}{dt} = 0$, the equations 1 and 2 became equation 3 and 4 respectively.

$$[E2Crimson]_{SS} = \frac{k}{k_d}\left(b_1 + \frac{[C_{22}H_{22}N_2O_7]^{n_1}}{K_1^{n_1} + [C_{22}H_{22}N_2O_7]^{n_1}}\right)\left(b_2 + \frac{[C_{10}H_{15}NO_4]^{n_2}}{K_2^{n_2} + [C_{10}H_{15}NO_4]^{n_2}}\right) \text{ equation 3}$$

$$[E2Crimson]_{SS} = \frac{k'}{k'_d}\left(b_3 + \frac{[C_9H_{18}O_5S]^{n_3}}{K_3^{n_3} + [C_9H_{18}O_5S]^{n_3}}\right)\left(b_4 + \frac{[C_{10}H_{15}NO_4]^{n_4}}{K_4^{n_4} + [C_{10}H_{15}NO_4]^{n_4}}\right) \text{ equation 4}$$

Now, when we kept one inducer concentration contant at its staturatated concentration, the equation 3 reduced in the following equations,

$$[E2Crimson]_{SS} = constant\_1 \left( b_1 + \frac{[C_{22}H_{22}N_2O_7]^{n_1}}{K_1^{n_1} + [C_{22}H_{22}N_2O_7]^{n_1}} \right) \quad \textbf{equation 5}$$

when $[C_{10}H_{15}NO_4]$ was kept constant and

$$[E2Crimson]_{SS} = constant\_2 \left( b_2 + \frac{[C_{10}H_{15}NO_4]^{n_2}}{K_2^{n_2} + [C_{10}H_{15}NO_4]^{n_2}} \right) \textbf{ equation 6}$$

when $[C_{22}H_{22}N_2O_7]$ was kept constant. Similarly, equation 4 was reduced to equation 7 and 8.

$$[E2Crimson]_{SS} = constant\_3 \left( b_3 + \frac{[C_9H_{18}O_5S]^{n_3}}{K_3^{n_3} + [C_9H_{18}O_5S]^{n_3}} \right) \textbf{ equation 7}$$

when $[C_{10}H_{15}NO_4]$ is constant and

$$[E2Crimson]_{SS} = constant\_4 \left( b_4 + \frac{[C_{10}H_{15}NO_4]^{n_4}}{K_4^{n_4} + [C_{10}H_{15}NO_4]^{n_4}} \right) \textbf{ equation 8}$$

when $[C_9H_{18}O_5S]$ is kept constant.

Next we fitted the dose response curves in figure 2 C, D, G, and H with equations 5, 6, 7, and 8 respectively. In each case we found that the Hill coefficients, which measures the sensitivity[28] of the output protein expression as a function of input chemical concentration, were greater than 1 (Fig 2 D, E, I, J), suggesting ultrasensitive or digital like behavior[28] for both the AND gates A and B. The all parameter fitting values were shown in table S3.

**Construction and characterization of six 3 input AND gates**

Next, we expanded these two 2-input genetic AND circuits and build six different 3-input molecular genetic AND gates (Fig 3A). First, we placed mVenus, cerulean, and E-2crimson genes under $P_{AAH}$ synthetic promoter in AND gate A and inserted separately in E-coli DH5αZ1 cells to make cell based molecular genetic AND gates, named as AND gate 1, 2, and 3 respectively (Fig. 3A). Similarly, we placed mVenus, Tdtomato, and E2-crimson genes under $P_{IAH}$ synthetic promoter in AND gate B in *E-coli* DH5αZ1 cells to make AND gates 4,5, and 6 respectively (Fig. 3A). Next, we grew each cell set (AND gate) separately in minimal media with glucose ($C_6H_{12}O_6$) as a sole carbon source with all possible combinations of inducers $C_9H_{18}O_5S$, $C_{22}H_{22}N_2O_7$, and $C_{12}H_{19}NO_4$ and experimentally measured the fluorescence protein expression. The results for each AND gate gene circuit showed correct match with the expected input-output relation tables (Fig. 3B).

Further, we tested the interference, if any of the 4$^{th}$ chemicals for all the six 3-input AND circuits by characterizing the behavior of the circuits in the presence of all four chemicals. The results suggested there was none (Fig. 3B). Here, glucose worked as sole carbon source and was responsible for the cell growth, without which, the cell would not grow and the circuits would not give any fluorescence. Now, we measured the fluorescence after culturing the cells in minimal media with the inducers and in the presence (1) and absence (0) of glucose and the results showed no fluorescence from the cell systems (Fig S1) without glucose. This matched well with the respective truth tables (Fig. 3B). Thus, for all the six AND gates, glucose worked as one of the inputs.

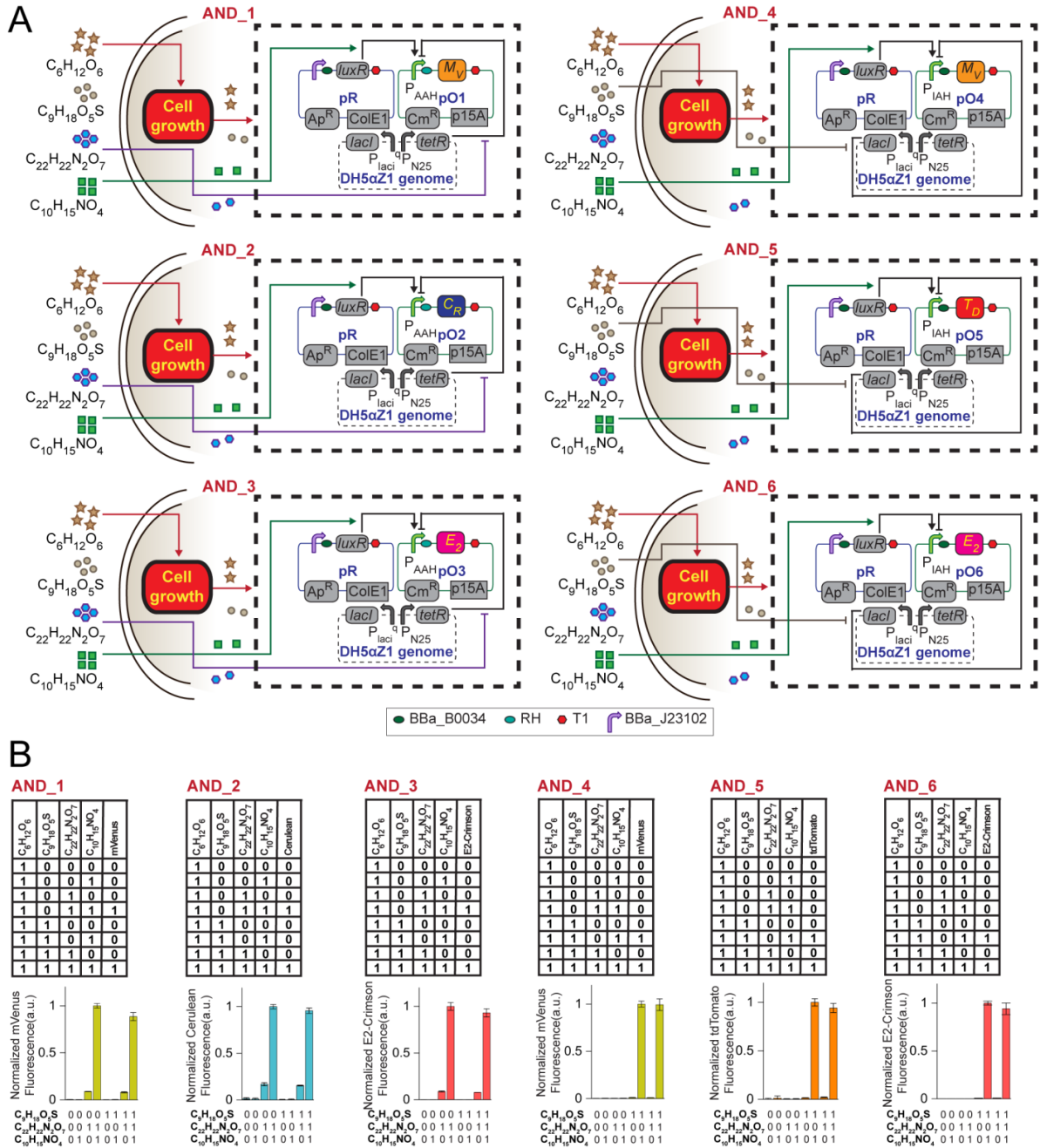

**Figure 3: Details of AND gate systems AND_1-6. A)** Synthetic genetic circuit designs of AND gates 1-6. **B)** Truth table and experimental characterization of AND gates 1-6 in presence of $C_6H_{12}O_6$ (D-glucose).

**Solving the maze with engineered cell populations**

Next, to complete the whole integrated cellular circuits (Fig. 1C), we mixed the six cell sets containing six AND gates and co-cultured them together in minimal media. This mixed populations of cells worked as a biological computer, which would be able to solve any 2X2 maze problem generated chemically using four molecules $C_6H_{12}O_6$, $C_9H_{18}O_5S$, $C_{22}H_{22}N_2O_7$, and $C_{12}H_{19}NO_4$ according to the input matrices (Fig 1). We added these four chemicals to the cell populations in minimal media following the input problem matrix (Fig. 1A) and allowed the culture to grow for 48 hours. Next, we collected the cells washed them PBS twice and re-suspended in PBS and measured the fluorescence outputs from the cell culture using a confocal fluorescence microscope. We observed that glowed (expressed florescence protein) only if a solution existed. Further, the mapping of the glowing bacteria on the maze according to the designated spaces showed the solution path (Fig 4). It was visualized that there were solutions existed for three problems (Fig 4) and absence of solutions were observed in other thirteen problems, where no fluorescence was observed (Fig 4). Those behaviors were appropriately matched with the corresponding input matrices -output matrices relation map (Fig. 1). The merged pictures for each case were shown in figure S2.

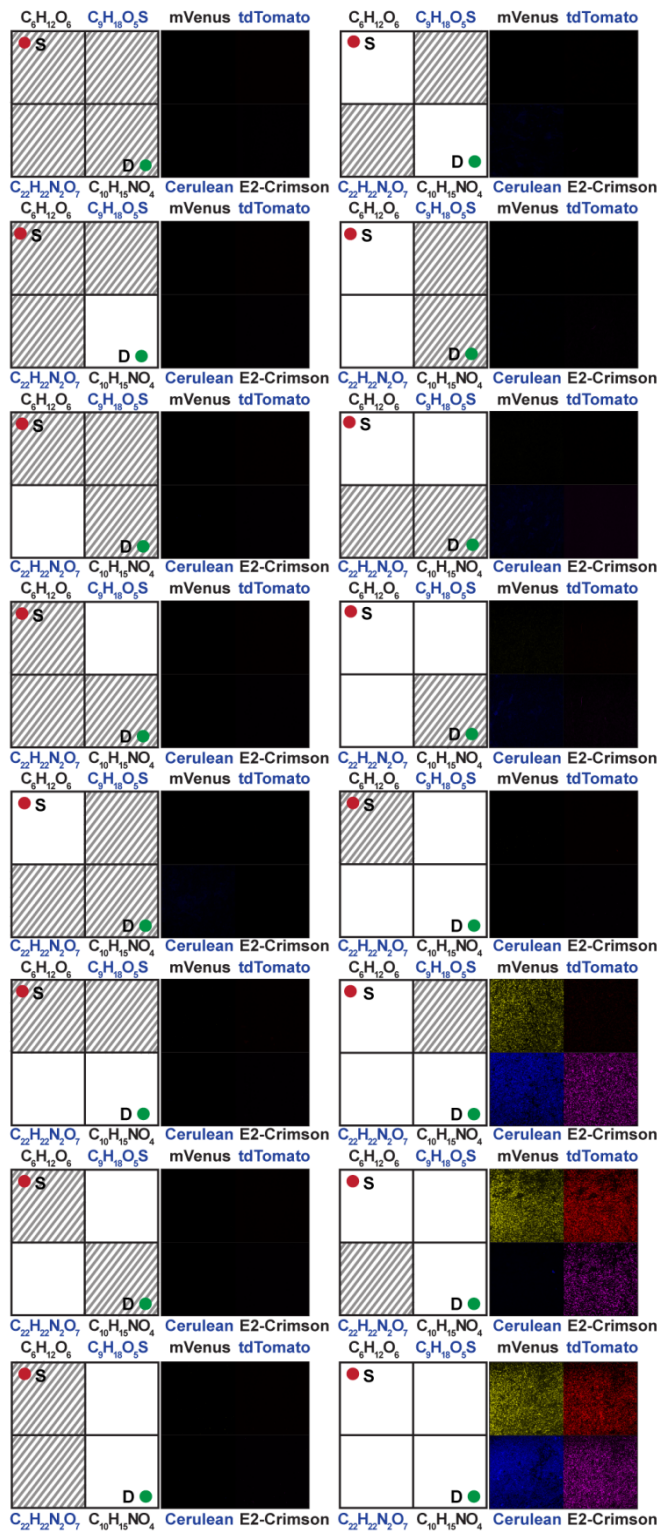

**Figure 4: Cellular computation solutions of the 2X2 chemical maze problem observed under the fluorescence microscope.** Images corresponding to four fluorescent protein channels (mVenus, tdTomato, Cerulean and E2-Crimson) are displayed as the solution of

each problem. Mazes with available solutions appeared fluorescent and mazes with no solution appeared dark. The corresponding maze problems were also shown in grid format.

In this work we created a set of engineered bacteria, which in a mixed population could solve all plausible 2X2 chemically generated maze problems. Unlike the other cellular maze solving, where the problem is limited by a physical structure and its solution depends on some chemotaxic gradient[10-12], this work presented a new and abstract cellular computational way to look at the maze problems using engineered bacteria. Here, we generated abstract maze problems by assigning chemicals at each position of the maze, and its presence or absence defined the 'obstruction' or 'open space'. Thus, a mathematical question was posed through chemicals. We created six sets of molecular engineered bacteria, which in a mixed population worked as a 'solver'. The engineered bacteria consisted of genetic logic circuits, through which the solution algorithm has been implemented. The 'solver' processed the chemical information and produced a solution by expressing fluorescence proteins and the solution path(s) was visualized by glowing bacteria. If there was no solution, the system became dark.

This maze-solving task was an example of biological distributed computing[29], where the total computational task was distributed among six different cell populations with six different genetic circuits. In theory, the general strategy used here can be scaled up to more bigger maze problems by increasing the number of inputs, outputs, and genetic circuits. Further, in this work the maze generation and its solutions are coupled together. As a result, unlike the other cell based maze solutions, our synthetic genetic system not only solved the maze problems but also demonstrated the number of solvable and unsolvable problems.

Interestingly, mathematically such a calculation was not straightforward[30]. Generating such abstract mathematical mazes using chemicals and solving with engineered bacteria beyond physical maze might inspire new possibilities in cellular computing and information processing. It may have significance in cellular cryptography, biocomputation, and synthetic biology.

**Methods**

**Plasmids, Bacterial cell strain, cell culture**

The genetic circuits were created based on the pZ expression system[26], and *Escherichia coli* DH5αZ1 strain was used as the chassis for the AND gate systems. Details of the AND gate systems, plasmids, bacterial strain, promoters and RBSs used in this study are listed in supplementary tables S1,2,4. The final concentration of the antibiotics in the LB Agar, Miller (Himedia) plates, LB broth, Miller (Himedia) and M9 minimal media was 100 μg/ml and 34 μg/ml for ampicillin (Himedia) and chloramphenicol (Himedia) respectively. $C_9H_{18}O_5S$ was purchased from Himedia, whereas $C_6H_{12}O_6$, $C_{22}H_{22}N_2O_7$ and $C_{10}H_{15}NO_4$ were purchased by Sigma Aldrich. The composition of M9 minimal media was as follows: 1X M9 salt solution, 5 g/L ammonium chloride, 2 g/L D-glucose (Molecular formula: $C_6H_{12}O_6$), 1 mM magnesium sulphate, 0.3 mM calcium chloride, 1 μg/mL thiamine-HCL and 1X trace element solution. pH of the 10X stock of M9 salt solution (Mixture of 422.65 mM sodium phosphate dibasic, 220.44 mM potassium phosphate monobasic and 85.56 mM sodium chloride) was adjusted to ~7.3. 100X trace element solution was made by mixing 13.4 mM ethylenediamine tetraacetic acid (EDTA, pH 7.5), 3.1 mM ferrous chloride hexadydrate, 0.076 mM cupric chloride dihydrate, 0.62 mM Zinc chloride, 0.042 mM Cobalt Chloride dihydrate, 0.162 mM boric acid and 0.0081 mM manganese chloride tetrahydrate.

**Characterization and dose response experiments of the AND gate systems**

For characterization of AND gates A and B in LB media, and AND gates 1-6 in M9 minimal media, their respective overnight cultures in LB media were diluted 100 times in fresh media with dual antibiotics ampicillin and chloramphenicol as well as appropriate combinations of chemical inputs and again grown for 16 hours (For LB media) and 48 hours (For M9 minimal media) at 37° C, ~ 250 rpm. Cells were then washed and re-suspended in phosphate buffered saline (PBS; pH ~7.3), diluted with PBS to reach OD value at 600 nm around 0.8, loaded into a black flat-bottom 96-well plate (Greiner) and both OD600 value and fluorescence were measured in Synergy HTX Multi-Mode reader (Biotek Instruments, USA) using appropriate combinations of excitation and emission filters for Cerulean, mVenus, tdTomato and E2-Crimson (440/30 nm, 500/27 nm, 540/35 nm and 610/10 nm band-pass excitation filters respectively; 485/20 nm, 540/25 nm, 590/20 nm and 645/10 nm band-pass emission filters respectively). Input '0' logic states for D-glucose, IPTG, aTc and 3-oxo-C6-AHL were 0 units, whereas, their input '1' logic states were 2 g/L, 10 mM, 200 mg/mL and 1 μM respectively. The cell palettes corresponding to the overnight cultures of AND_1-6 in LB media were washed with M9 glucose$^+$ minimal media before re-inoculating them into fresh M9 minimal media. For dose response experiments, the concentration of one inducer was varied keeping the other one constant at its saturated concentration. In all cases, data from 3 independent bacterial colonies per experimental set was taken and DH5αZ1 cells carrying no plasmids were considered as the control set.

**Observing biological maze problems using fluorescence microscopy**

The cell pellets from the separate single colony-overnight cultures of the six AND gate systems (AND_1-6) in LB media were washed twice with and resuspended in M9 glucose$^+$ minimal media. Next, an equal volume of each of those resuspended cells was mixed into M9 minimal media supplemented with antibiotics and the chemical inputs with appropriate combinations. All the samples were then incubated for 48 hours, washed at least three times with PBS, resuspended in fresh PBS and observed under 60 X water immersion in Nikon AIR si confocal microscope supplied with the resonant scanner and coherent CUBE diode laser system. The sample fields were subjected to excitation of the fluorescent proteins by the sequential firing of 4 laser channels (405 nm, 488 nm, 561 nm and 640 nm Lasers) and their emissions were captured by 4 different emission filters (BP 450/50 nm, BP 525/50 nm, BP 585/65 nm and BP 700/75 nm filters). The brightness and contrasts of the microscopic images were adjusted using Fiji software.

**Data analysis and fitting**

For the charecterization and dose response of basic AND gates A and B, the absolute fluorescence value for each bacterial colony of the experimental set (s) as well as the DH5αZ1 no plasmid set was normalized to the number of cells by dividing it by its respective OD value at 600 nm:

$$\text{Normalized fluorescence of a single bacterial colony} = \left[\frac{\text{Absolute fluorescence}}{\text{OD600}}\right] \quad \ldots$$

**Equation 9**

Next, autofluorescence was removed by subtracting the average normalized fluorescence of the DH5αZ1 no plasmid set from the normalized fluorescence of individual colonies of the experimental sets. Finally, the average normalized fluorescence without autofluorescence was calculated for each experimental set:

$$\text{Normalized fluorescence} = \left[\frac{\sum \text{Normalized fluorescence without autofluorescence}}{\text{Number of bacterial colonies}}\right] \pm \text{S.D} \quad \ldots$$

**Equation 10**

For the charecterization of the AND gates 1-6 in minimal media, average fluorescence was obtained as

$$\text{Average fluorescence} = \left[\frac{\sum \text{Absolute fluorescence}}{\text{number of bacterial colonies}}\right] \pm \text{S.D}$$

... **Equation 11**

All data analysis and curve fitting were performed using Origin2018 (OriginLab Corporation, USA) software. The curve fitting of the dose response curves were performed with the appropriate equations using the built-in Levenberg−Marquardt algorithm, a damped least-squares (DLS) method.

**Acknowledgements**

This work was financially supported by SINP intramural funding (Department of Atomic Energy, Govt. of India), SERB (CRG/20lB/00l394), and Ramanujan Fellowship (DST), Govt. of India

**Author Contributions**

SB conceived and designed the study. KS and DB performed all the experiments. KS and SB designed the experiments, analyzed and interpreted the data, and wrote the paper.

**Competing Interests statement**

We have no competing interest.

**Supplementary Information**

**Engineered Bacteria Computationally Solve Chemically Generated 2X2 Maze Problems**

Kathakali Sarkar, Deepro Bonnerjee, Sangram Bagh[*]

Biophysics and Structural Genomics Division, Saha Institute of Nuclear Physics, Homi Bhabha National Institute (HBNI), Block A/F, Sector-I, Bidhannagar, Kolkata 700064 INDIA

[*] Corresponding author E-mail: sangram.bagh@saha.ac.in

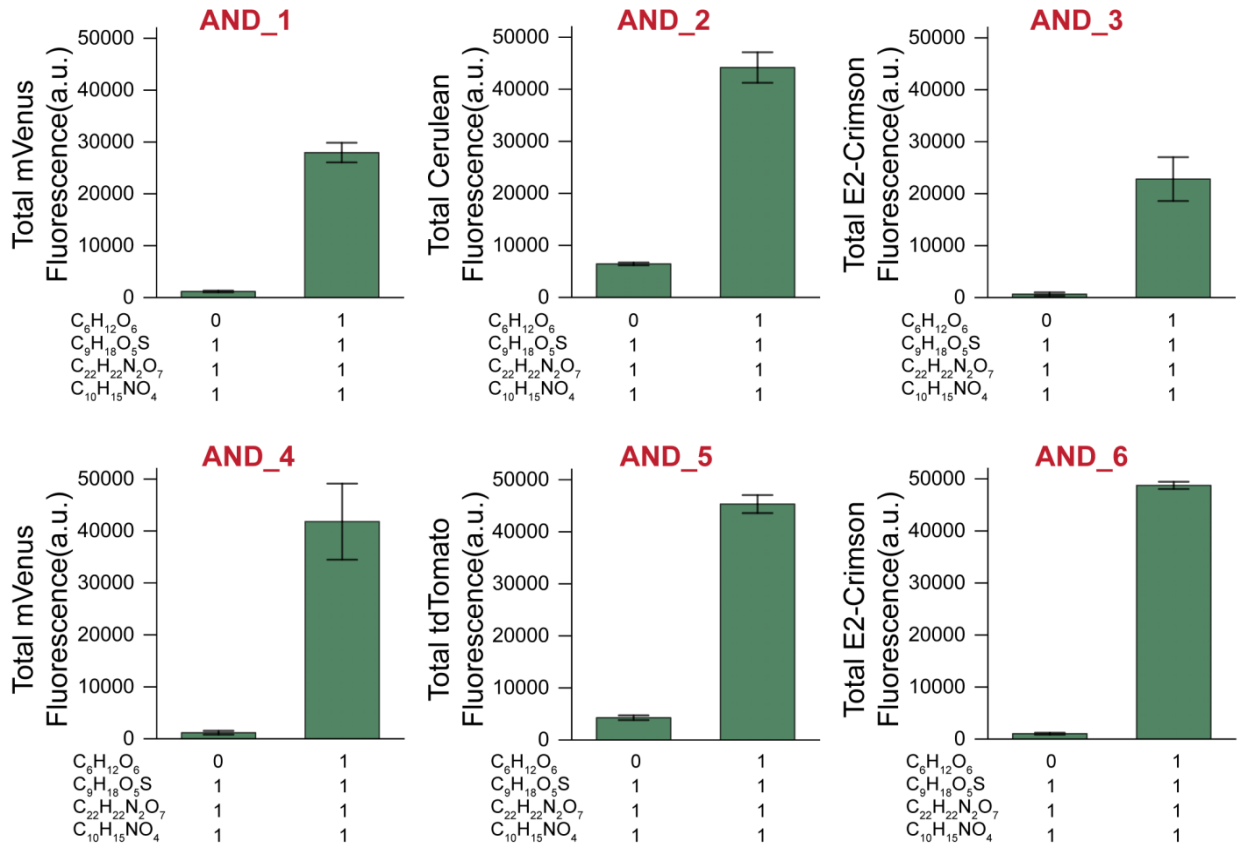

**Figure S1:** Experimental characterization of AND gates (AND_1-6) in M9 minimal media supplemented with or without D-glucose ($C_6H_{12}O_6$) while the rest of the input chemicals are present.

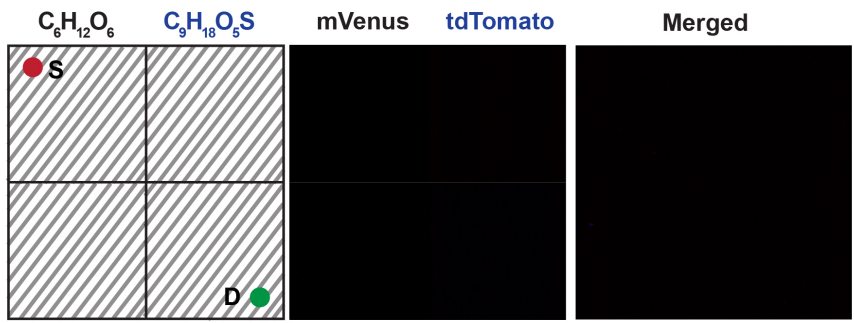
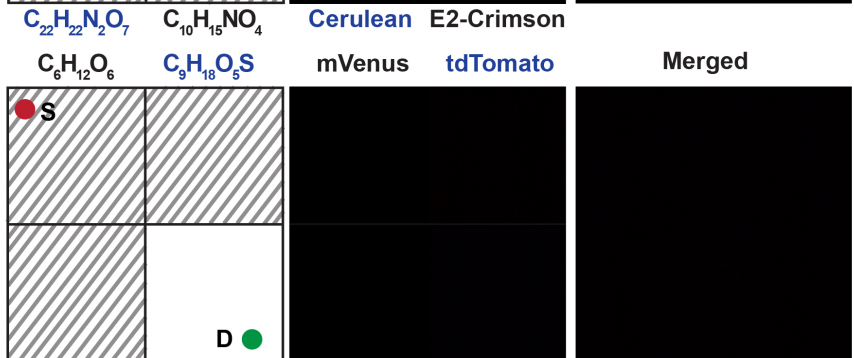
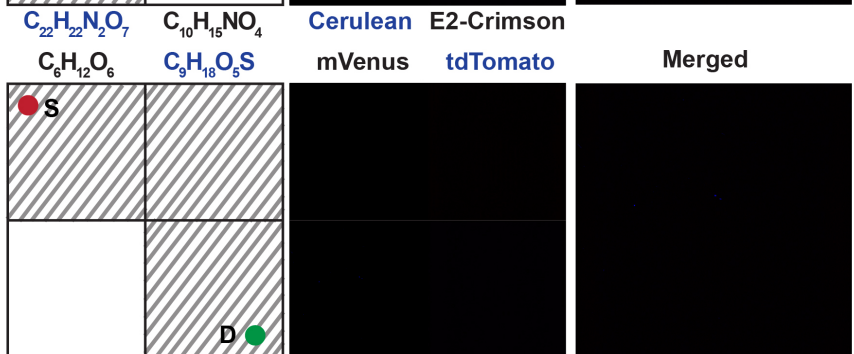
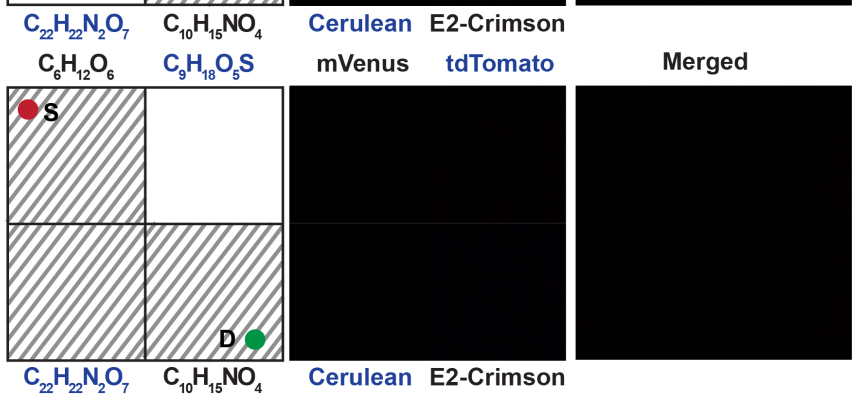

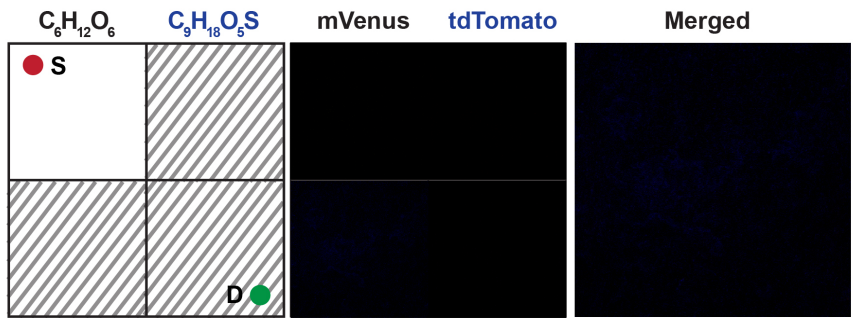
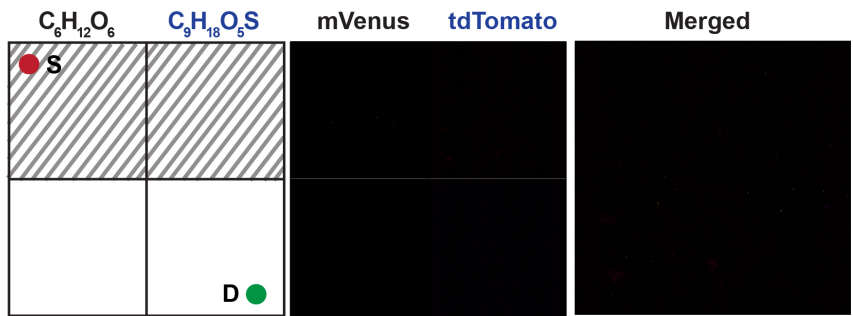
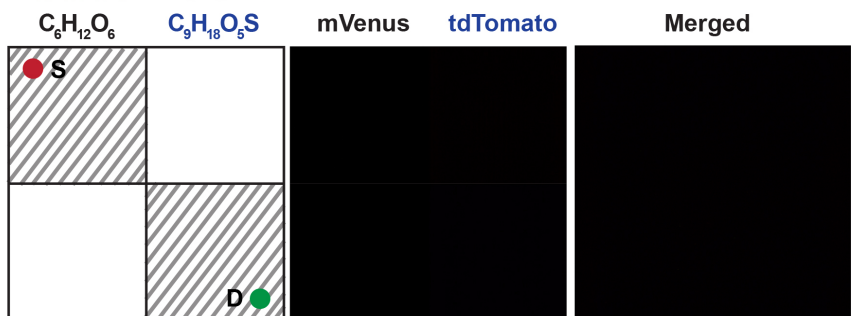
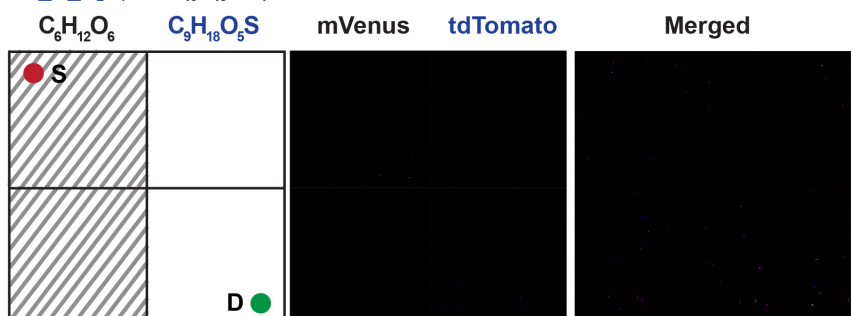

| $C_6H_{12}O_6$ | $C_9H_{18}O_5S$ | $C_{22}H_{22}N_2O_7$ | $C_{10}H_{15}NO_4$ | mVenus | tdTomato | Cerulean | E2-Crimson |
|---|---|---|---|---|---|---|---|
| 0 | 0 | 0 | 0 | 0 | 0 | 0 | 0 |
| 0 | 0 | 0 | 1 | 0 | 0 | 0 | 0 |
| 0 | 0 | 1 | 0 | 0 | 0 | 0 | 0 |
| 0 | 1 | 0 | 0 | 0 | 0 | 0 | 0 |
| 1 | 0 | 0 | 0 | 0 | 0 | 0 | 0 |
| 0 | 0 | 1 | 1 | 0 | 0 | 0 | 0 |
| 0 | 1 | 1 | 0 | 0 | 0 | 0 | 0 |
| 0 | 1 | 0 | 1 | 0 | 0 | 0 | 0 |
| 1 | 0 | 0 | 1 | 0 | 0 | 0 | 0 |
| 1 | 0 | 1 | 0 | 0 | 0 | 0 | 0 |
| 1 | 1 | 0 | 0 | 0 | 0 | 0 | 0 |
| 1 | 1 | 1 | 0 | 0 | 0 | 0 | 0 |
| 0 | 1 | 1 | 1 | 0 | 0 | 0 | 0 |
| 1 | 0 | 1 | 1 | 1 | 0 | 1 | 1 |
| 1 | 1 | 0 | 1 | 1 | 1 | 0 | 1 |
| 1 | 1 | 1 | 1 | 1 | 1 | 1 | 1 |

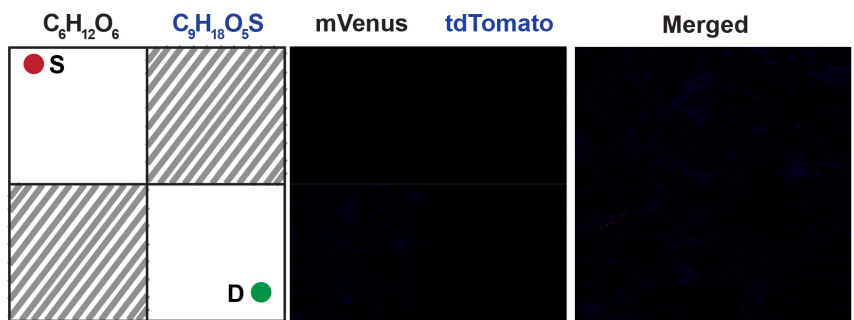
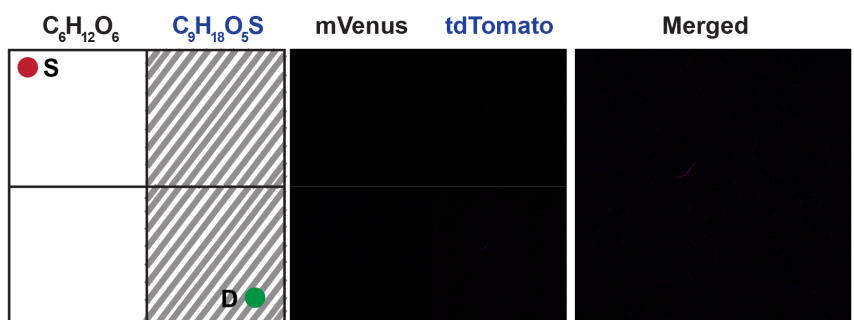
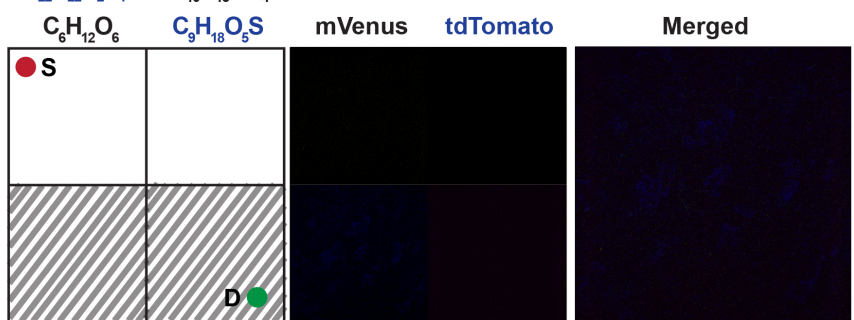
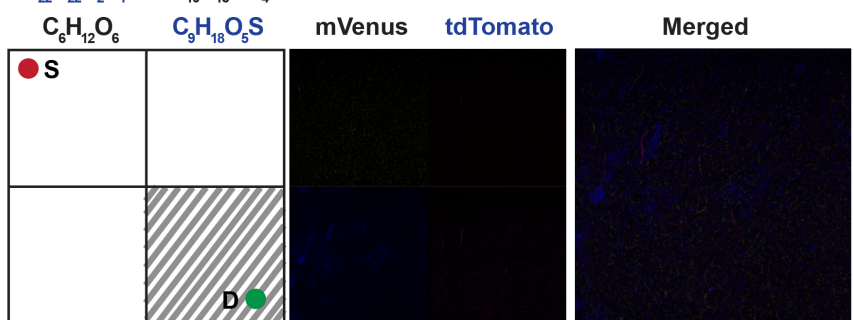

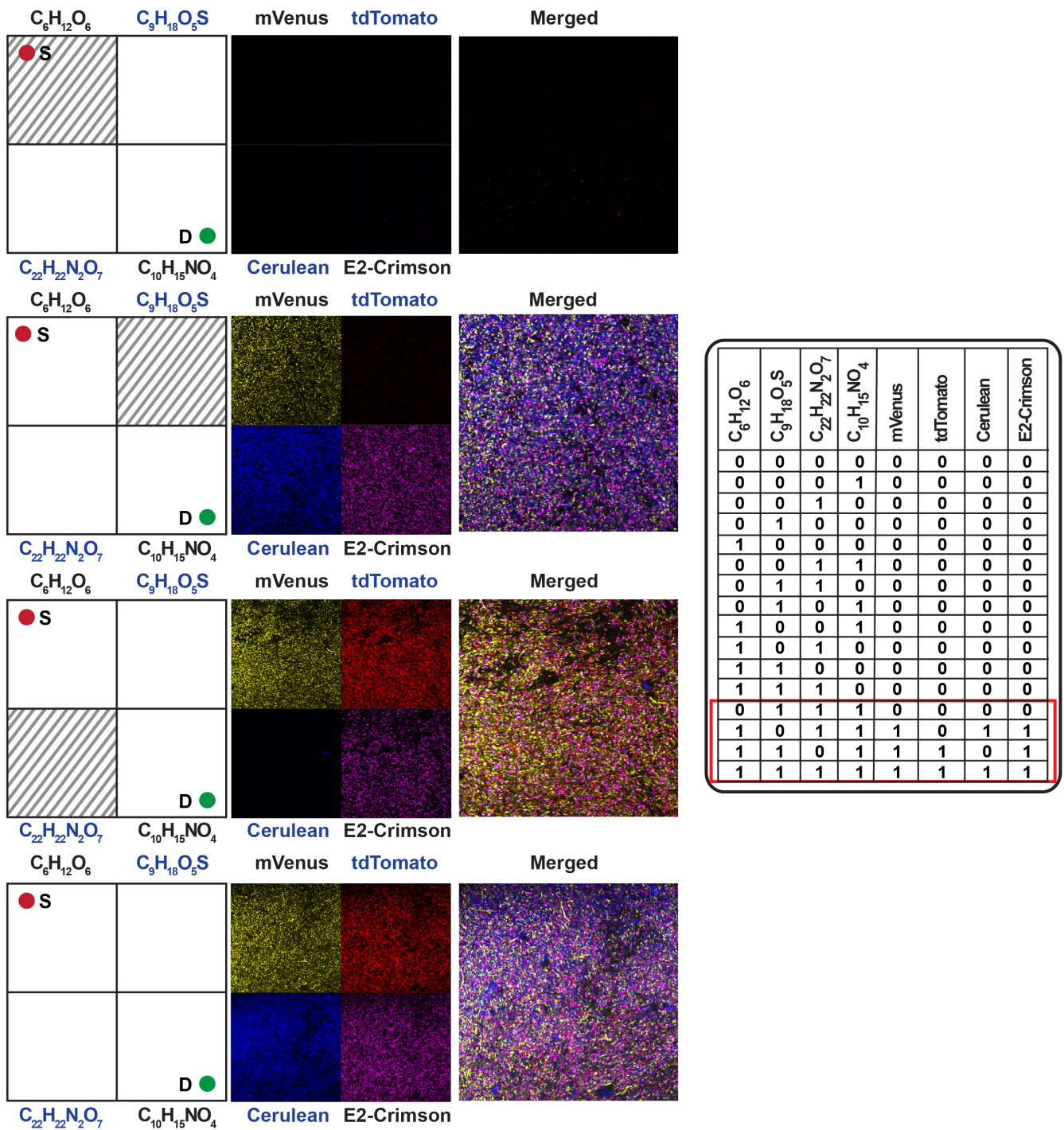

**Figure S2: Microscopic observation of the 2X2 chemical maze problem.** Each of the 16 possible problems and the experimental observation of its corresponding solution are shown. Images were taken for 4 channels corresponding to 4 fluorescent proteins (mVenus, tdTomato, Cerulean and E2-Crimson) for the same sample field and the merged image of all channels is also shown. Data corresponding to the input-output relation is pointed in the truth table with the red box.

**Table S1: List of Promoters and RBSs.** *lac*O1, *tet*O2 and Lux box sites are colored in red, brown, and green respectively. Transcription start site is shown in bold. -10 and -35 hexamers are underlined. Each promoter is flanked by *Xho*I and *EcoR*I restriction sites and RBS is flanked by *EcoR*I and *Kpn*I restriction sites. Restriction sites are marked in italics.

| Name | Sequence (5´ – 3´) | Source |
|---|---|---|
| P$_{AAH}$ | *CTCGAG*ACCTGTAGGATCGTACAGGTTTACGTCCCTATCAGTGATAGAGTATAGTCGAATA**A**ATCCCTATCAGTGATAGAGA*GAATTC* | [1] |
| P$_{IAH}$ | *CTCGAG*ACCTGTAGGATCGTACAGGTTTACGTTTGTGAGCGGATAACAATATAGTCGAATA**A**ATTGTGAGCGGATAACAATT*GAATTC* | [1] |
| BBa_J23102 | *CTCGAG*TTGACAGCTAGCTCAGTCCTAGGTACTGTGCTAGC*GAATTC* | [2] |
| BBa_B0034 | *GAATTC*ATTAAAGAGGAGAAA*GGTACC* | [3] |
| RH | *GAATTC*ATTGGAGAGGAGTCC*GGTACC* | [1] |

**Table S2: List of bacterial strain and plasmids.**

| Plasmid name | Description | Ori | Antibiotic selection | Source |
|---|---|---|---|---|
| *E. coli* DH5αZ1 | The chassis | - | - | Prof. David McMillen |
| pOR-EGFP-12 | Source of ColE1 ori | ColE1 | Amp | Prof. David McMillen |
| pOR-Luc-31 | Source of p15A ori | p15A | Cm | Prof. David McMillen |
| pmVenus-C1 | Source of mVenus gene | pUC | Kan | Clontech |
| pmCerulean-C1 | Source of Cerulean gene | pUC | Kan | Clontech |
| ptdTomato | Source of tdTomato gene | pUC | Amp | Clontech |
| pUCP20T-E2Crimson (Plasmid#78473) | Source of E2-Crimson gene | pBR322 | Amp | Addgene |
| pBW313lux-hrpR (Plasmid#61436) | Source of LuxR gene | p15A | Kan | Addgene |
| pP$_{AAH}$C3RBSHEGFP(R) | EGFP gene with RBS RH under P$_{AAH}$ promoter | p15A | Cm | [1] |
| pO1 | mVenus gene with RBS RH under P$_{AAH}$ promoter | p15A | Cm | This study |
| pO2 | Cerulean gene with RBS RH under P$_{AAH}$ promoter | p15A | Cm | This study |
| pO3 | E2-Crimson gene with RBS RH under P$_{AAH}$ promoter | p15A | Cm | This study |
| pP$_{IAH}$C3EGFP(R) | EGFP gene with RBS BBa_B0034 under P$_{IAH}$ promoter | p15A | Cm | [1] |
| pO4 | mVenus gene with RBS BBa_B0034 under P$_{IAH}$ promoter | p15A | Cm | This study |
| pO5 | tdTomato gene with RBS BBa_B0034 under P$_{IAH}$ promoter | p15A | Cm | This study |
| pO6 | E2-Crimson gene with RBS BBa_B0034 under P$_{IAH}$ promoter | p15A | Cm | This study |
| pR | LuxR gene with RBS BBa_B0034 under BBa_J23102 promoter | ColE1 | Amp | This study |

**Table S3: Parameter values obtained from curve fitting.**

| AND gate | Dose response of chemical input | Parameter values with standard deviations | | | | | | | |
|---|---|---|---|---|---|---|---|---|---|
| | | c | | b | | k | | n | |
| | | Value | S.D. | Value | S.D. | Value | S.D. | Value | S.D. |
| AND gate A | $C_{22}H_{22}N_2O_7$ | 0.95353 | 0.03454 | 0.0006 | 0.000425 | 0.00007 | 0.000004 | 3.21486 | 0.2276 |
| | $C_{10}H_{15}NO_4$ | 0.96529 | 0.0569 | 0.07558 | 0.00514 | 0.0001 | 0.000022 | 1.14353 | 0.9362 |
| AND gate B | $C_9H_{18}O_5S$ | 0.90582 | 0.00829 | 0.13058 | 0.00729 | 0.00674 | 0.000936 | 1.29826 | 0.19071 |
| | $C_{10}H_{15}NO_4$ | 1.00000 | 0.000000 | 0.0178 | 0.00125 | 0.00002 | 0.000008 | 3.35887 | 1.29137 |

**Table S4: List of AND gate systems.**

| AND gate system | Output plasmid | | | | | Regulator plasmid | | | | |
|---|---|---|---|---|---|---|---|---|---|---|
| | Promoter - output protein cassette | RBS | Ori | Antibiotic selection | Plasmid name | Promoter-regulator cassette | RBS | Ori | Antibiotic selection | Plasmid name |
| AND_1 | $P_{AAH}$-mVenus | RH | P15A | Cm | pO1 | BBa_J23102-LuxR | BBa_B0034 | ColE1 | Amp | pR |
| AND_2 | $P_{AAH}$-Cerulean | RH | P15A | Cm | pO2 | | | | | |
| AND_3 | $P_{AAH}$-E2-Crimson | RH | P15A | Cm | pO3 | | | | | |
| AND_4 | $P_{IAH}$-mVenus | BBa_B0034 | P15A | Cm | pO4 | | | | | |
| AND_5 | $P_{IAH}$-tdTomato | BBa_B0034 | P15A | Cm | pO5 | | | | | |
| AND_6 | $P_{IAH}$-E2-Crimson | BBa_B0034 | P15A | Cm | pO6 | | | | | |